\documentclass[12pt]{book}

\usepackage[dvips]{graphicx,color}
\usepackage{makeidx,tsukuba}
\makeauthorindex
\makeindex

\begin{document}

\BookTitle{\itshape The 28th International Cosmic Ray Conference}
\CopyRight{\copyright 2003 by Universal Academy Press, Inc.}
%\tableofcontents
\pagenumbering{arabic}

\chapter{%   %%%%%%%%% <===== TITLE of the contribution
%%%%%%%%%%% The first letter of each word should be capital letter.
Hourly Spectral Variability of Mrk~421}

\author{%
%
% You can include as many co-authors as you wish, unless
% the title/author information fits within 1 page.
%
F.~Krennrich$^{1,2}$, I.H.~Bond, P.J.~Boyle, S.M.~Bradbury, J.H.~Buckley,
D.~Carter-Lewis, O.~Celik, W.~Cui, M.~Daniel, M.~D'Vali,
I.de~la~Calle~Perez, C.~Duke, A.~Falcone, D.J.~Fegan, S.J.~Fegan,
J.P.~Finley, L.F.~Fortson, J.~Gaidos, S.~Gammell, K.~Gibbs,
G.H.~Gillanders, J.~Grube, J.~Hall, T.A.~Hall, D.~Hanna, A.M.~Hillas,
J.~Holder, D.~Horan, A.~Jarvis, M.~Jordan, G.E.~Kenny, M.~Kertzman,
D.~Kieda, J.~Kildea, J.~Knapp, K.~Kosack, H.~Krawczynski,
M.J.~Lang, S.~LeBohec, E.~Linton, J.~Lloyd-Evans, A.~Milovanovic,
P.~Moriarty, D.~Muller, T.~Nagai, S.~Nolan, R.A.~Ong, R.~Pallassini,
D.~Petry, B.~Power-Mooney, J.~Quinn, M.~Quinn, K.~Ragan, P.~Rebillot,
P.T.~Reynolds, H.J.~Rose, M.~Schroedter, G.~Sembroski, S.P.~Swordy,
A.~Syson, V.V.~Vassiliev, S.P.~Wakely, G.~Walker, T.C.~Weekes,
J.~Zweerink \\
{\it  (1) Physics \& Astronomy Department, Iowa State University, Ames, IA, 50011
}
{\it  (2) The VERITAS Collaboration--see S.P.Wakely's paper} ``The VERITAS
Prototype'' {\it from these proceedings for affiliations}
}%% end of author

\section*{Abstract}
Mrk~421 is the first TeV blazar found to exhibit significant spectral variability 
during strong flaring activity, showing hardening of the TeV spectrum in high 
emission states.   
Mrk~421 is also known to exhibit flux variability on time scales as short as 15 minutes.  
In this paper we present studies of hourly spectral variability of Mrk~421 in 2001 
using data from the Whipple Observatory 10~m gamma-ray telescope.

\section{Introduction}
The AGN phenomenon is generally associated with strongly varying
fluxes at all wavelengths on time scales of hours to several months and years. 
At TeV gamma-ray energies, flux variability as short as 15 minutes 
have been observed for Mrk~421 (Gaidos et  al., 1996).
Flux  variations are useful information providing the grounds for dynamical
tests for particle acceleration and/or emission models, e.g., 
constraining cooling time scales and physical parameters of the source such
as the magnetic field and size of the emission region.

To further the understanding of non-thermal emission from AGN jets,
multiwavelength observations are pursued by  measuring the spectrum 
over many orders of magnitude in energy and its variations.  Correlations 
between X-ray and TeV emission were found in various flaring  episodes   
for Mrk~421 (Buckley et al. 1996; Maraschi et al. 1999; Jordan et al. 2001).
A next step in testing jet models can be provided by measurements of the
spectral variations of an X-ray  (Fossati et al. 2000) and contemporaneous 
TeV gamma-ray flare. 
Results of spectral variability of Mrk~421 were shown previously (Aharonian et al. 2002; 
Krennrich et al. 2002) by the  HEGRA and VERITAS collaborations.    In this paper
we discuss spectral variability of flares occuring on two nights in 
March 2001.

\section{Results:}
Unusually intense and lasting flaring activity of Mrk~421 in 2001 gave
excellent statistics  and detailed features of its energy spectrum 
have been derived:    Mrk~421 exhibits a curved spectrum that can be 
described by a power law with an exponential cutoff around 4~TeV 
(Krennrich et al. 2001; Aharonian et al. 2002).
 
\begin{figure}[t]
  \begin{center}
    \includegraphics[height=21pc]{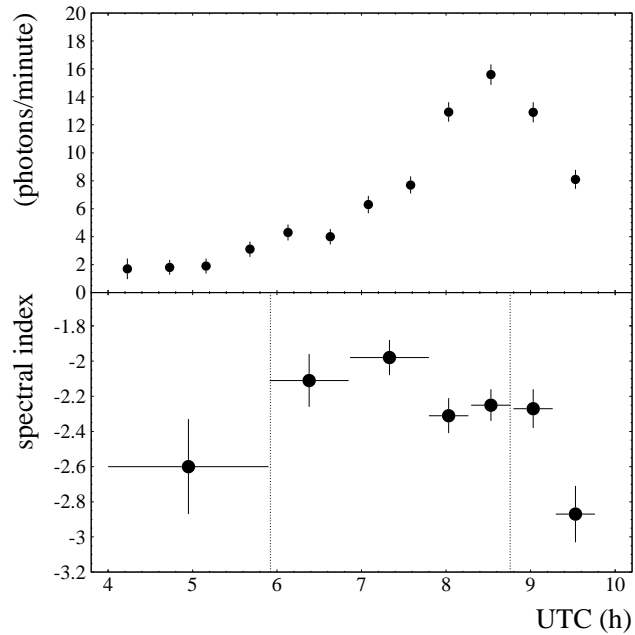}
  \end{center}
  \vspace{-0.5pc}
  \caption{The lightcurve of a flare of Mrk~421 on March 19 2001 is shown together with the
             differential spectral index.  The probability that the spectral variability 
              is statistical is $\rm 3.1 \times 10^{-4}$ (significance $\rm \approx  
              3.6  \sigma$).  The lower plot shows a division (dotted line) into three 
               different stages of the flare into preflare, rising flare and postflare. }
\end{figure}
\noindent The energy spectral index varies as a function of flux when averaging
over several months of data, but no  evidence 
for variability in the cutoff energy was found when analyzing the entire 2001 data set. 
 Therefore,  in the following analysis  of spectral variability on hourly time scales, 
since it is based on spectra from small subsets of the 2001 data with less statistics,  
we fit the  data using a power  law  with  a fixed exponential cutoff energy  of 4.3 TeV 
(see also Krennrich et al. 2002).

\begin{figure}[t]
  \begin{center}
    \includegraphics[height=21pc]{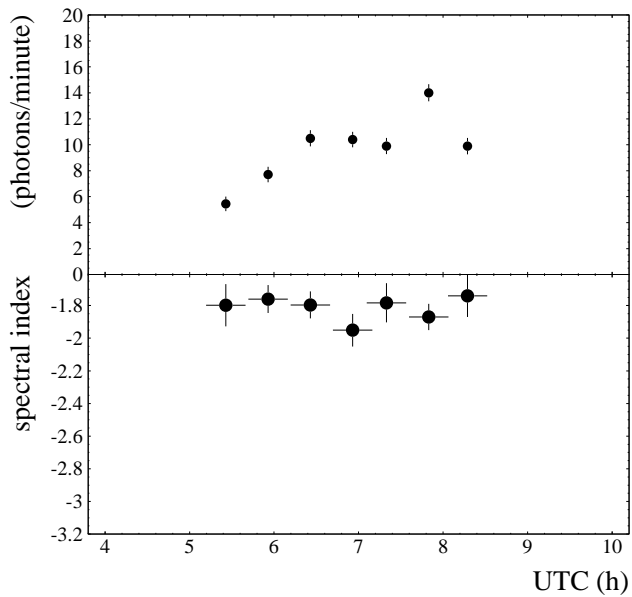}
  \end{center}
  \vspace{-0.5pc}
  \caption{The lightcurve of a flare of Mrk~421 on March 25 2001 is shown together with the
             differential spectral index.  The spectral index is constant despite the fact that
               the flux changes by a factor of 2.5. }
\end{figure}
Figure 1 shows the gamma-ray lightcurve for Mrk~421 on March~19 2001 (Jordan et al. 2001;
Buckley 2001) for which the 
TeV observations provide almost complete sampling of the rise and decay of a flare. 
The gamma-ray  rate ranges from a moderate  level ($\rm \approx$~1~Crab) 
up to an eight  fold flux increase in less  than 4.5 hours.   
The lower part of figure 1 shows the  spectral index measured at various times during
the flare. 
The hypothesis that the spectral index is constant during the flare  has a   chance 
probability of $\rm 3.1 \times 10^{-4}$, suggesting spectral variability during this
outburst of gamma-ray emission with a significance of $\rm \approx 3.6  \sigma$.
   
When dividing the spectral index measurements into three different episodes,  
preflare,  rising flare and postflare, the hardest spectral  index coincides with 
the rise of the flare, whereas during the preflare and the 
postflare the spectrum appears to be softer.
A similar correlation between hardening/softening of the energy spectrum and flux 
for Mrk~421 has been reported by the HEGRA collaboration  (Aharonian et al. 2002) 
for the nights of March 21/22 and 22/23 2001.

\noindent However, observations of flaring activity during other nights suggest that the relation
between spectral index and flux variability is more complex than suggested by data
from March 19, 21/22 and 22/23 2001.  
Representative of the complexity of spectral characteristics during flares of other 
nights is a flare on March 25 2001 measured with the Whipple 10~m telescope.     
Figure 2 shows its lightcurve with significant flux variations and a flux
increase by a factor of 2.5 within $\rm \approx 2$~hours. 
 
The spectral index during this flare is exceptionally hard 
($\rm \alpha \: = \:  1.82 \: \pm 0.04$) without any indications
of spectral variability (hypothesis of constant spectral index has a probability of
78.5\%.).  Although substantial flux variations occured no significant spectral variations
are detected, despite good statistics. 
It is noteworthy that the data from this flare do not include a preflare episode 
with fluxes of the level of 1~Crab and do not include the postflare era.   This may
indicate the spectral variations on hourly time scales may be  dependent on the
absolute flux.

%   Although substantial flux variations occured no significant spectral variations
% are detected, despite the fact that statistics are fairly rich for all the data points.
% It is noteworthy that the flux values for this flare are all above approximately above
% 3~Crab, hence all data points show the source in  a high flaring state.

\section{Conclusions}
In this paper we present  two  lightcurves for hourly flaring activity of Mrk~421 
together with spectral index variations.  During a flare on March 19 2001 significant
spectral variability  is suggested by the data showing  a soft spectral index 
in the preflare and postflare phase and a hard spectral index
during the rising phase of the flare.   In contrast no spectral variations are observed 
for the night of March 25 despite significant flux variations.  

One-component models that describe gamma-ray flares solely by an increase in the maximum energy
of the radiating particle distribution are ruled out by this observation.  
A more comprehensive analysis involving a large number of flares is beyond the scope
of this paper but  will be presented at the conference.

\section{Acknowledgements}
We acknowledge the technical assistance of E. Roache and J. Melnick.  This research is
supported by grants from the U.S. Department of Energy, the Enterprise Ireland and by
PPARC in the UK.

% These data demonstrate the 

\section{References}
\re
Aharonian, F.A., et al. 1999, A\&A, 351, 330
\re
Aharonian, F.A., et al. 2002, A\&A, 384, L23
\re
Buckley, J.H., et al. 1996, ApJ, 472, L9
\re
Buckley, J.H. 2001, AIP Conf. Proceedings 587, Gamma 2001, ed. S. Ritz, 
N. Gehrels, C.R. Shrader, p 235 
\re
Fossati, G., et al. 2000, ApJ, 541, 166
\re
Gaidos, J.A., et al. 1996, Nature, 383, 319
\re
Jordan, M., et al. 2001, in Proc. 27th ICRC (Hamburg), OG2.3, 2691
\re
Krennrich, F., et al. 2001, ApJ, 560, L45
\re
Krennrich, F., et al.  2002, ApJ, 575, L9
\re
Maraschi, L., et al. 1999, ApJ, 526, L81
\re
\endofpaper
\end{document}